\documentclass[pre,floats,twocolumn,showpacs,superscriptaddress]{revtex4}

\usepackage{epsfig}
\usepackage{latexsym,amsmath}

\begin{document}

\title{Community detection in complex networks using Extremal Optimization}

\author{Jordi Duch}

\affiliation{Departament d'Enginyeria Inform{\`a}tica i
Matem{\`a}tiques,
  Universitat Rovira i Virgili, 43007 Tarragona, Spain}

\author{Alex Arenas}

\affiliation{Departament d'Enginyeria Inform{\`a}tica i
Matem{\`a}tiques,
  Universitat Rovira i Virgili, 43007 Tarragona, Spain}

\date{\today}

\begin{abstract}
  We propose a novel method to find the community structure in complex
  networks based on an extremal optimization of the value of modularity. The method outperforms the optimal modularity found by the existing algorithms in the literature. We present the results of the algorithm for computer simulated and real networks and compare them with other approaches. The efficiency and accuracy of the method make it feasible to be used for the accurate identification of community structure in large complex networks.
\end{abstract}

\maketitle

The description of the structure of complex networks has been one of the
focus of attention of the physicist's community in the recent
years. The levels of description range from the microscopic
(degree, clustering coefficient, centrality measures, etc., of
individual nodes) to the macroscopic description in terms of
statistical properties of the whole network (degree distribution,
total clustering coefficient, degree-degree correlations, etc.)
\cite{s-ecn-01,ab-smcn-02,dm-eon-02,n-sfcn-03}. Between 
these two extremes there is a "mesoscopic" description of networks 
that tries to explain its community structure. The general notion of
community structure in complex networks was first pointed out in the physics literature by
Girvan and Newman \cite{gn-cssbn-03}, and refers to the fact that
nodes in many real networks appear to group in subgraphs in which
the density of internal connections is larger than the connections with
the rest of nodes in the network.

The community structure has been empirically found in many real
technological, biological and social networks \cite{Eckmann02,esms-meei-03,shbp-hhj-03, addgg-casn-03, Newman04}
and its emergence seems to be at the heart of the network
formation process \cite{gddga-scso-03}.


The existing methods intended to devise the community structure in complex networks have been recently reviewed in \cite{Newman04}. All these methods require a definition of
community that imposes the limit up to which a group should be
considered a community. However, the concept of community itself
is qualitative: nodes must be more connected within its community
than with the rest of the network, and its quantification is still
a subject of debate. Some quantitative definitions that came from
sociology have been used in recent studies \cite{rcclp-dicn-04},
but in general, the physics community has widely accepted a recent
measure for the community structure based on the concept of
modularity $Q$ introduced by Newman and Girvan \cite{ng-fecsn-04}:

\begin{equation}
    Q=\sum_{r} ( e_{rr} - a_{r}^{2} )
\label{Q}
\end{equation}

\noindent where $e_{rr}$ are the fraction of links that connect two
nodes inside the community $r$,  $a_{r}$ the
fraction of links that have one or both vertices inside of the
community $r$, and the sum extends to all communities $r$ in a given
network. Note that this measure provides a way to
determine if a certain mesoscopic description of the graph in
terms of communities is more or less accurate. The larger the
values of $Q$ the most accurate a partition into communities is.

The search for the optimal (largest) modularity value is a NP-hard
problem due to the fact that the space of possible partitions grows faster than any power of the system size. For this reason, a
heuristic search strategy is mandatory to restrict the search
space while preserving the optimization goal \cite{n-fadcsn-04}.
Indeed, it is possible to relate the current optimization problem
for $Q$ with classical problems in statistical physics, e.g. the
spin glass problem of finding the ground state energy
\cite{sk-smsg-75}, where algorithms inspired in natural
optimization processes as simulated annealing \cite{kgv-osa-83}
and genetic algorithms \cite{g-gasoml-89} have been successfully
used.

In this Letter, we propose a new divisive algorithm that optimizes
the modularity $Q$ using an heuristic search based on the Extremal
Optimization (EO) algorithm proposed by Boettcher and Percus
\cite{bp-oed-01,bp-eogp-01}. This algorithm is inspired in turn in the
evolution model of Bak-Sneppen \cite{bs-pecsme-93}, and basically
operates optimizing a global variable by improving extremal local
variables that involve co-evolutionary avalanches. The performance
of EO algorithms have been shown to overcome the efficiency of
classical simulated annealing and genetic algorithms providing
competitive accuracy \cite{bp-nwo-00}.

In our case, the global variable to optimize is $Q$ as defined in
eq.(\ref{Q}). Thus, the definition of the local variables used in the extremal optimization problem should be related to the contribution of individual nodes $i$ to the
summation in eq.(\ref{Q}) given a certain partition into communities

\begin{equation}
    q_{i}=\kappa_{r(i)} - k_{i} a_{r(i)}
\label{q_j}
\end{equation}

\noindent where $\kappa_{r(i)} $ is the number of links that a node
$i$ belonging to a community $r$ has with nodes into the same community, and $k_i$ is the degree of node $i$.
Note that $Q=\frac{1}{2L}\sum_{i}q_i$ where $i$ refers to all nodes
in the network given a certain partition into communities and $L$ is the
total number of links in the network. Eq.(\ref{q_j}) provides a measure that depends on the node degree,
and its normalization involve all the links in the network after
summation. Re-scaling the local variable $q_{i}$ by the degree of
node $i$ we obtain a proper definition for the contribution of
node $i$ to the modularity, relative to its own degree and
normalized in the interval [-1,1].

\begin{equation}
    \lambda_{i} = \frac{q_{i}}{k_i} = \frac{\kappa_{r(i)}}{k_i} - a_{r(i)}
\label{lambda_j}
\end{equation}

Keeping in mind this definition of $\lambda_{i}$ we can
compare the relative contribution of individual nodes to the
community structure. We will consider $\lambda_{i}$ as the local
variable involved in the extremal optimization process that
characterizes an individual node, from now on we will refer to
$\lambda_{i}$ as the fitness of node $i$ using the common jargon in
extremal optimization problems.

The heuristic search we propose to find the optimal modularity
value evolves as follows: 
\begin{itemize}

\item Initially, we split the nodes of the
whole graph in two random partitions having the same number of
nodes each one. This splitting creates an initial communities division, where communities are understood as connected components in each partition. 

\item At each time step, the system self-organizes by moving the node with the lower fitness (extremal) from one partition to the other. In principle, each movement implies the recalculation of the fitness of many nodes because the right hand side of equation (\ref{lambda_j}) involves the pseudo-global magnitude $a_{r(i)}$.

 \item The process is repeated until an "optimal state" with a maximum value of Q is reached. After that, we delete all the links
between both partitions and proceed recursively with every resultant
connected component. The process finishes when
the modularity $Q$ could not be improved \footnote{The value of Q always refers to the whole network i.e. is the sum over all the communities. At a certain moment more subdivisions into communities will necessarily decrease Q because the limit of decomposition is a community per node whose value of Q is negative.}.

\end{itemize}

Note that this process is not a bipartitioning of the graph as known in computer science \cite{bp-eogp-01}, because: the number of nodes in each partition is dependent on the
evolution process and not restricted to be the same at the end of
the process; and more importantly, each partition could contain
different connected components (communities) that when the
partitions are disconnected result in several subgraphs. 

\begin{figure}[t]
  \epsfig{file=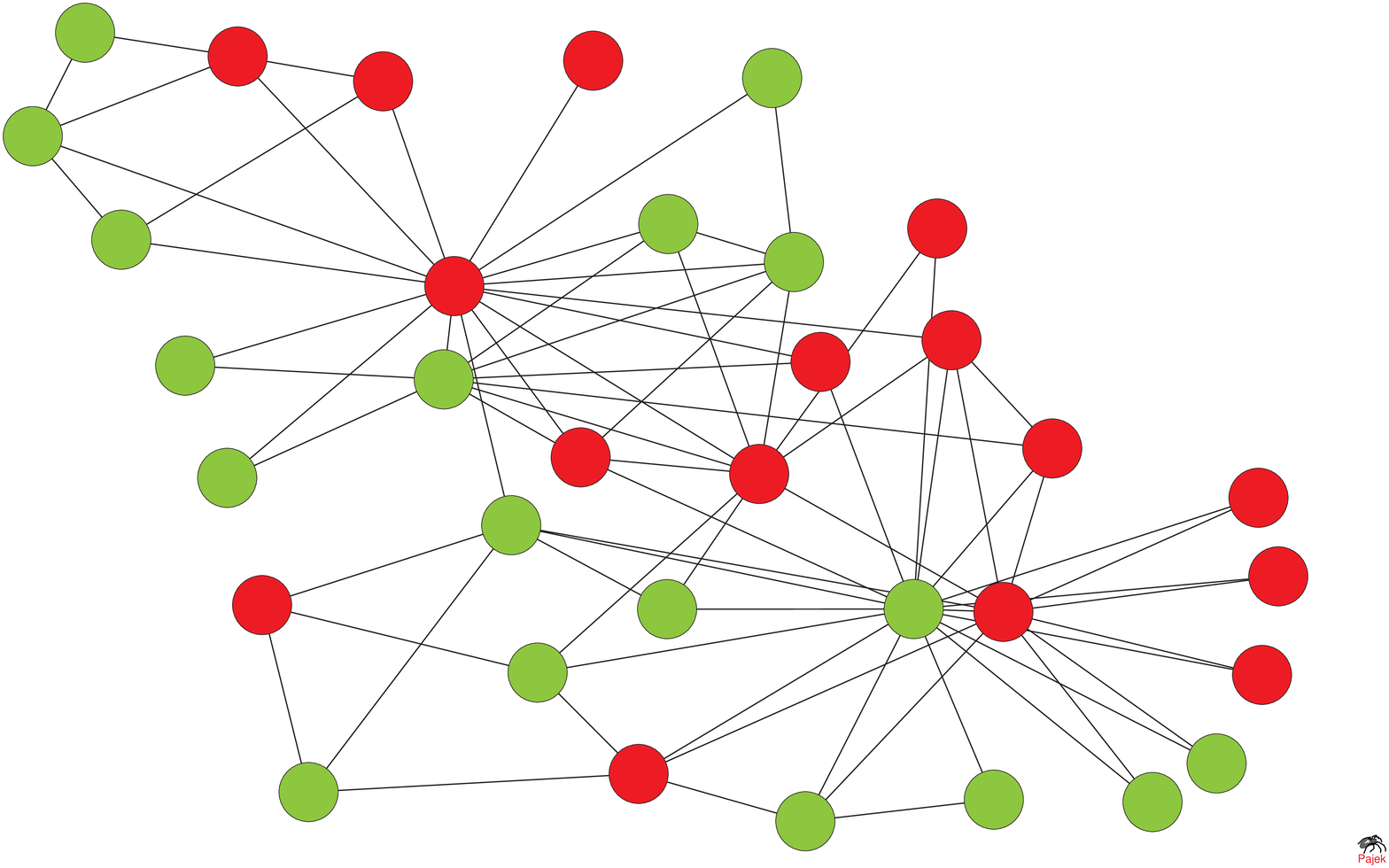, width=4cm}
  \epsfig{file=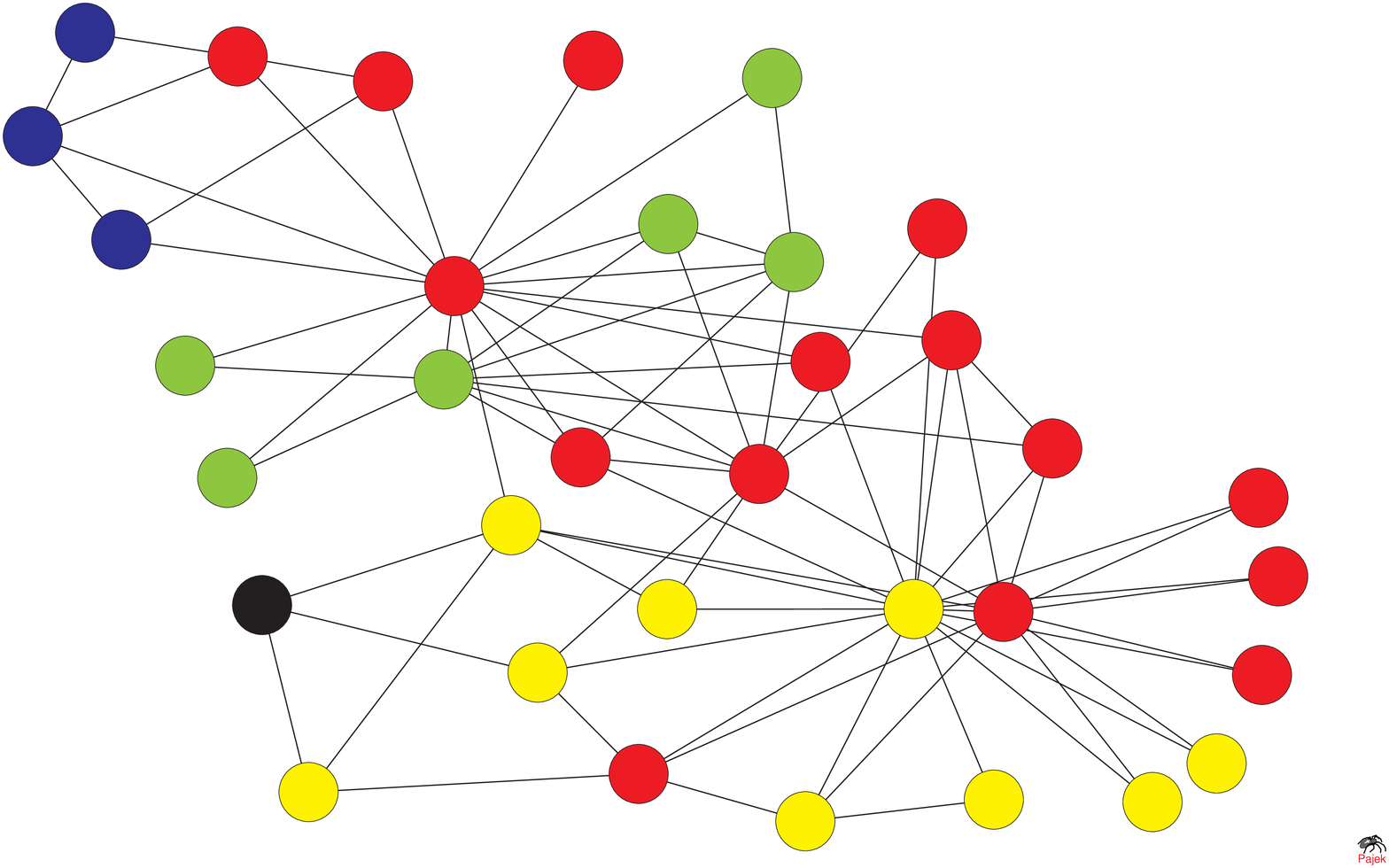, width=4cm}
  \caption{Left: Random initialization of the Zachary network into two partitions, red and green.
  Right: Five different communities identified as connected components in each partitions. Each color defines a different community.}
\label{initialization}
\end{figure}

\begin{figure}[b]
  \epsfig{file=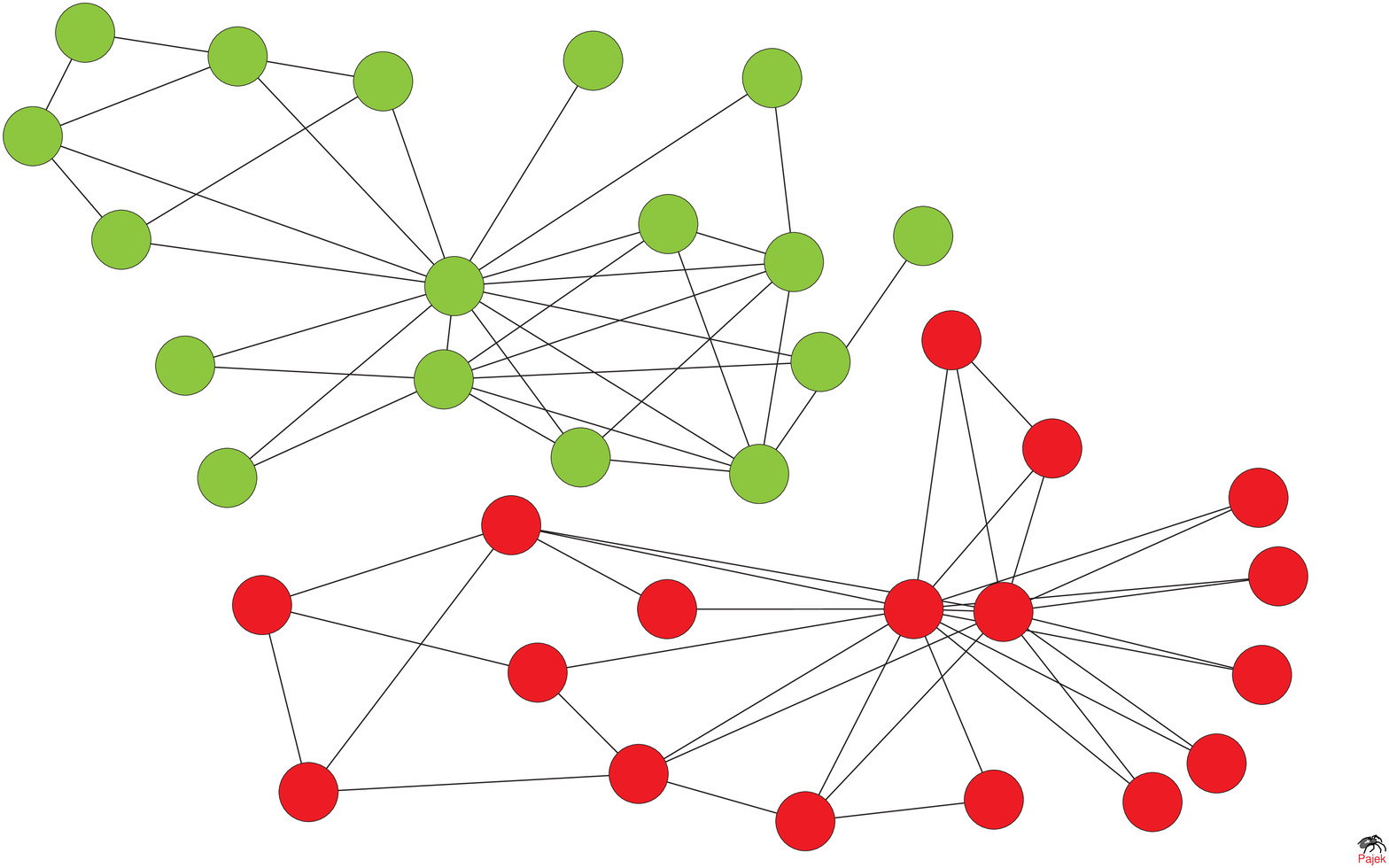, width=2.7cm}
  \epsfig{file=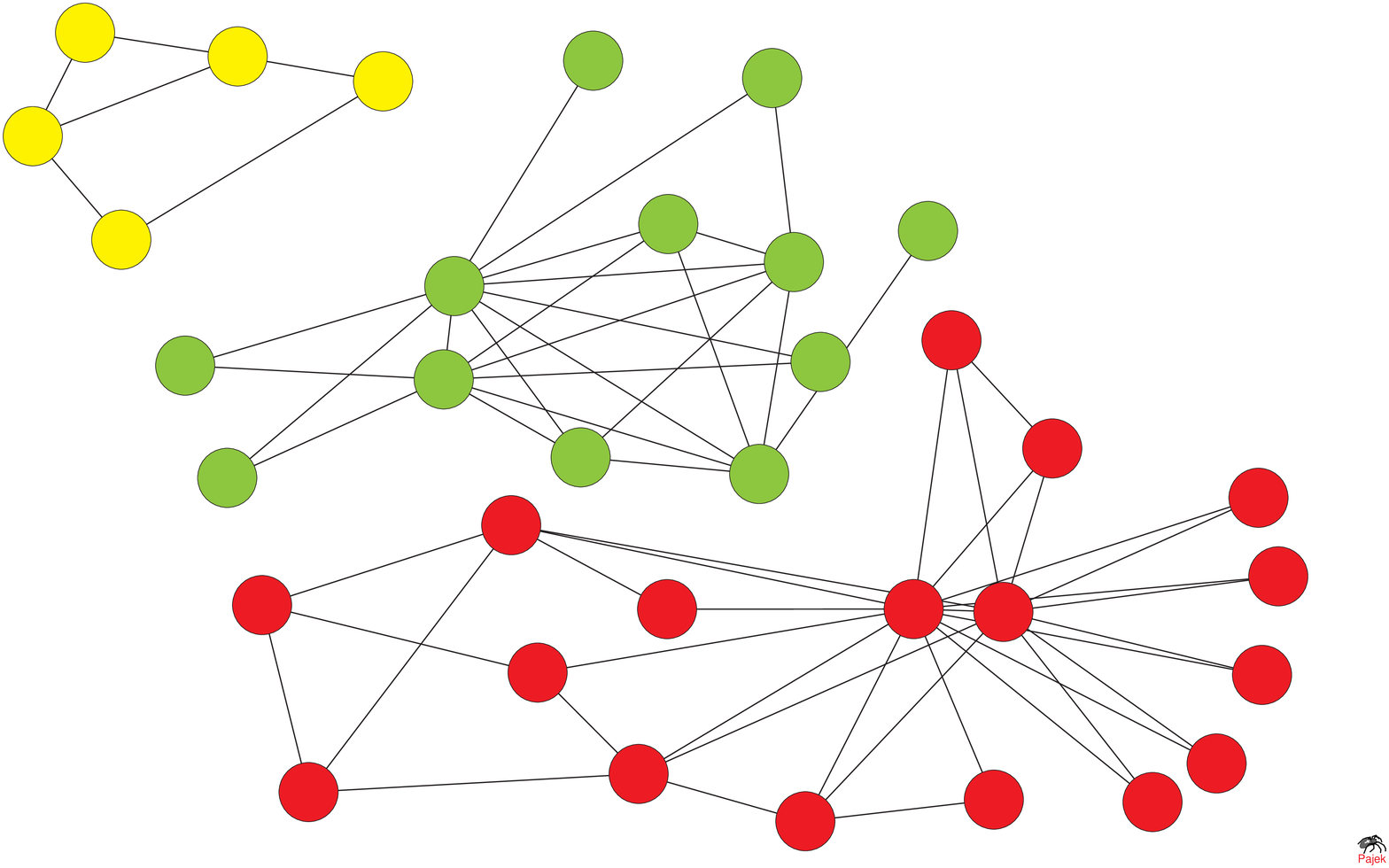, width=2.7cm}
  \epsfig{file=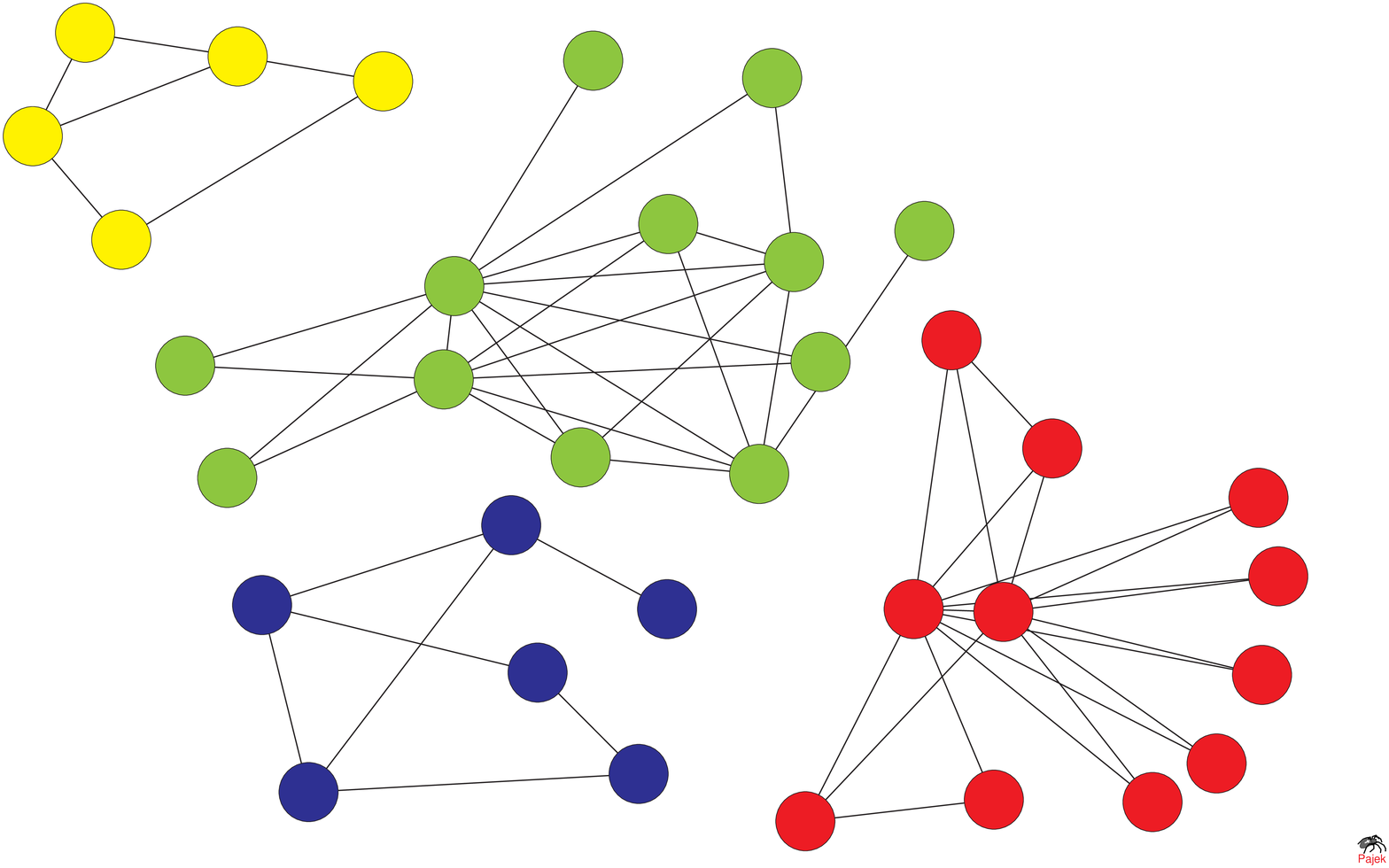, width=2.7cm}
  \epsfig{file=figure2d, width=8.1cm}

  \caption{Top: Network after edge removal at each recursive cut. Bottom: Evolution of the Q value in the at each step of the adaptation process. Separation bars indicate
  recursive divisions of the graph performed at maximum Q. }
\label{zachary}
\end{figure}

Let us illustrate the above mentioned heuristics in a simple case.
We will apply it to the well-know Zachary karate club
network \cite{z-ifmcfsg-77}. Initially we split the nodes in two
random partitions (see Fig.\ref{initialization} left). Note that the
number of initial communities (connected components in each
partition) in this case is five (see Fig.\ref{initialization}
right). After that, the self-organization process starts: the node with
the "worst fitness" is selected and moved from its partition to
the other partition, this movement provokes an avalanche of
changes in the fitness of the rest of nodes. We calculate the new
value for the modularity $Q$, and again repeat the process until
no changes could improve it (see Fig. \ref{zachary}).

The application of the algorithm to the Zachary network provides
the optimal modularity value after three recursive iterations. The
network is decomposed in four communities and the value for the
modularity is $0.419$, greater than the value $0.381$ reported by
Newman \cite{n-fadcsn-04}, the value $0.406$ reported by Reichardt
et al. \cite{rb-dfcsc-04} and the value $0.412$ reported by Donetti
et al. \cite{dm-dncns-04} using different optimization methods.

The extremal optimization (EO) approach presented here has several
technical implementation details, that are relevant for our purposes. 
In the original EO algorithm, the node selected is
always the node with the worst $\lambda_j$ value. This is a
deterministic and fast way to solve the problem, but it presents some
drawbacks: the final result strongly depends on the initialization
and there is no possibility to escape from local maxima. Instead,
we use a probabilistic selection called $\tau$-EO
\cite{bp-oed-01}, in which the nodes are ranked according to their
fitness values, and then the node of rank $r$ is selected
according to the following probability distribution:

\begin{equation}
    P(r) \propto r^{-\tau}
\label{ranking}
\end{equation}

This solution is less sensitive to different initializations and
allows to escape from local maxima. The exponent $\tau$ has been
tuned around the optimal values obtained for random networks of
size $N$ that approach the scaling $\tau \sim 1+1/ln(N)$ \cite{bp-oed-01}. The use
of this technique also implies the determination of the number of
self-organization steps $\alpha N$ needed to decide that the
maximum value has little chance to be improved. In practice, we keep track at each step of the last maximum value obatined for Q, if this maximum is not improved in  $\alpha N$ steps we stop the search. Usually $\alpha$ is
empirically determined balancing accuracy and efficiency in the
algorithm, we use $\alpha=1$ allowing as many steps as nodes to improve the current maximum value of Q. The computational cost involved in the whole process is
$O(N^{2}ln^{2}N)$ where a factor $NlnN$ is the cost associate to the ranking
process, however it can be substantially reduced using heap data
structures \cite{auh-dsa-83} for the ranking selection process up
to $O(N)$. The total cost of the algorithm can then be improved up to $O(N^{2}lnN)$.

\begin{figure}
  \epsfig{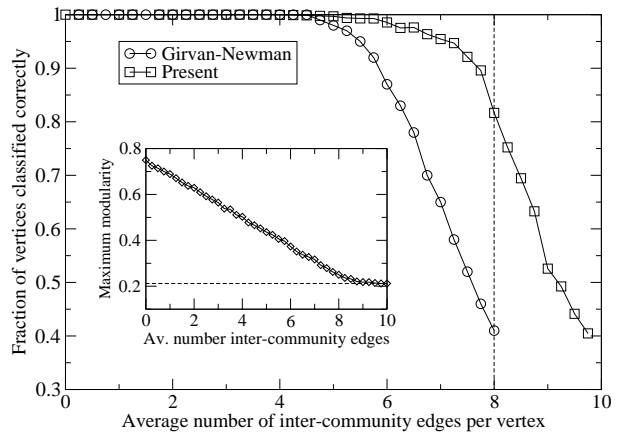}
  \caption{Fraction of nodes correctly classified using computer-generated graphs
  described in text. Each point is an average over 100 different networks.
  Inset: Average of the maximum modularity obtained in each case.}
\label{comp_zout}
\end{figure}

To test the performance of the algorithm we use first
computer-generated graphs with a known community structure
\cite{gn-cssbn-03}. These graphs have 128 vertices grouped in four
communities of 32 vertices. Each vertex has on average $z_{in}$
edges to vertices in the same community and $z_{out}$ edges to
vertices in other communities, keeping an average degree
$z_{in}+z_{out}=16$. We generate several graphs using $z_{out}$
values between 0 and 10, and compare the results of our algorithm
with those obtained using the heuristics proposed by Newman
\cite{n-fadcsn-04}. This shows the capabilities of each algorithm
identifying the communities when these are more fuzzy inside the
whole network. Using the Girvan-Newman algorithm, wich is the
reference algorithm for community identification, the communities
are well detected until values of $z_{out}=6$. In contrast, our
algorithm detects the communities up to $z_{out}=8$, where the
community structure still persist but is much more difficult to
reveal, see Fig.\ref{comp_zout} . In this particular case 50 percent of the links are within the community and 50 percent are links with nodes outside the community. This result that could seem contradictory is not. Note that the 50 per cent of links with nodes outside the community are equally distributed among the rest of communities, and then its contribution to the definition of community is deprived by the number of communities in the rest of the network, in our case three. For this reason it is expected to find community structure even in these cases.

For values higher than 8, the
average maximum modularity rapidly approach the limit $Q=0.208$
(see inset Fig.\ref{comp_zout}), the expected modularity for a
random network with the same number of links and nodes, as it has been shown 
in \cite{gsa-mfrgcn-04}.

We have also analyzed the community structure of several real
networks: the jazz musicians network \cite{gd-csj-03}, an
university e-mail network \cite{gddga-scso-03}, the $C. elegans$
metabolic network \cite{Jeong2000large-scale}, a network of users
of the PGP algorithm for secure information transactions
\cite{ggada-mmtn-02}, and finally the relations between authors
that shared a paper in cond-mat \cite{newman01a}.

\begin{table}[h]
\begin{tabular}{|c|c|c|c|c|c|}

\hline
Ê NetworkÊ &Ê SizeÊ &Ê $Q_{N}$Ê & $\#coms_N$ & $Q_{EO}$ & $\#coms_{EO}$Ê \\
\hline\hline
ÊÊÊ ZacharyÊÊÊÊ &ÊÊÊ 34 & 0.3810 &ÊÊ 2 & 0.4188 &ÊÊ 4 \\
\hline
ÊÊÊ JazzÊÊÊÊÊÊÊ &ÊÊ 198 & 0.4379 &ÊÊ 4 & 0.4452 &ÊÊ 4 \\
\hline
ÊÊÊ C. elegansÊ &ÊÊ 453 & 0.4001 &Ê 10 & 0.4342 &Ê 12 \\
\hline
ÊÊÊ E-mailÊÊÊÊÊ &Ê 1133 & 0.4796 &Ê 13 & 0.5738 &Ê 15 \\
\hline
ÊÊÊ PGPÊÊÊÊÊÊÊÊ & 10680 & 0.7329 &Ê 80 & 0.8459 & 365 \\
\hline
ÊÊÊ Cond-MatÊÊÊ & 27519 & 0.6683 & 302 & 0.6790 & 647 \\
\hline
\end{tabular}
Ê \caption{Maximum modularity obtained using the algorithm \cite{n-fadcsn-04} $Q_N$ and the extremal optimization algorithm $Q_{EO}$ for different complex networks. It is also included the number of communities found at the configuration with maximum modularity.}
\label{res}
\end{table}

In Table \ref{res} we present the results for the maximum
modularity achieved by our algorithm compared to the modularity
obtained using \cite{n-fadcsn-04}. The difference in
maximum modularity is up to 15\% depending on the network
considered. These differences result in a best determination of
the unknown community structure of the whole network. The partition into communities is clearly different  as shows the different number of communities found using both algorithms.

\begin{figure}[b]
 \epsfig{file=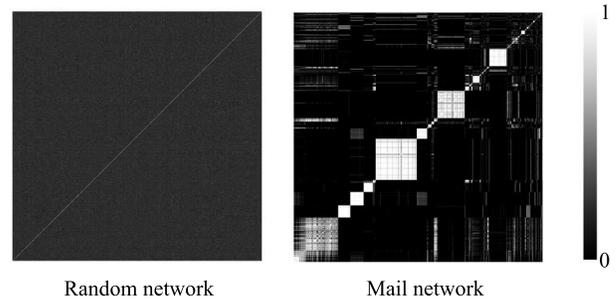, width=8cm}
  \caption{Fraction of nodes classified in the same partition over 100 realizations of the algorithm.
  The color of the position (i,j) corresponds to the fraction of times that nodes i and j belong to the same partition.}
\label{comprand}
\end{figure}

Note that since the core of the algorithm is stochastic, different runs
could yield in principle different partitions. We have performed
100 runs of the algorithm for the e-mail network and for a random
network with the same number of links and nodes to check the consistency 
of the proposed method. In Fig.
\ref{comprand} we present the results of the fraction of times a
couple of nodes are classified in the same partition. The community structure is clearly revealed for the e-mail network
while for the random network this structure is inexistent. Recently, Guimer\`a and Amaral have obtained similar results by applying simulated annealing to find the community structure in the context of metabolic networks \cite{ga-ccmn-05}.

Summarizing, we have presented an extremal optimization based 
algorithm that optimizes the modularity and allows an accurate identification of community structure in complex networks. The results outperform all previous algorithms existent in the literature.

We thank M. Bogu\~na , L. Danon, A. Diaz-Guilera and R. Guimer\`a for helpful comments and suggestions. We also thank M.E.J. Newman for providing us the cond-mat network. This work has been supported by DGES of the Spanish Government Grant No. BFM-2003-08258 and EC-FET Open Project No. IST-2001-33555.

\end{document}